# Distributed Voltage Regulation of Active Distribution System Based on Enhanced Multi-agent Deep Reinforcement Learning

Di Cao, *Student Member, IEEE,* Junbo Zhao, *Senior Member, IEEE,* Weihao Hu, *Senior Member, IEEE,* Fei Ding, *Senior Member, IEEE,* Qi Huang, *Senior Member, IEEE,* Zhe Chen, *Fellow, IEEE*

*Abstract*—This paper proposes a data-driven distributed voltage control approach based on the spectrum clustering and the enhanced multi-agent deep reinforcement learning (MADRL) algorithm. Via the unsupervised clustering, the whole distribution system can be decomposed into several sub-networks according to the voltage and reactive power sensitivity. Then, the distributed control problem of each sub-network is modeled as Markov games and solved by the enhanced MADRL algorithm, where each sub-network is modeled as an adaptive agent. Deep neural networks are used in each agent to approximate the policy function and the action value function. All agents are centrally trained to learn the optimal coordinated voltage regulation strategy while executed in a distributed manner to make decisions based on only local information. The proposed method can significantly reduce the requirements of communications and knowledge of system parameters. It also effectively deals with uncertainties and can provide online coordinated control based on the latest local information. Comparison results with other existing model-based and data-driven methods on IEEE 33-bus and 123-bus systems demonstrate the effectiveness and benefits of the proposed approach.

## I. Introduction

The utilization of renewable energy is of great significance to alleviate the current energy and environmental concerns [1]-[2]. However, due to the uncertainties and volatility characteristics of renewable energy, its higher integration brings numerous technical challenges to the distribution network (DN) operations. Among them, the overvoltage issue due to the inverse power flow and the voltage fluctuations caused by the volatility of renewable energy are paid increasing attention.

Various approaches have been proposed for DN voltage regulation. From the perspective of the control framework, they can be divided into three main categories: centralized, distributed autonomous, and distributed coordination control. The centralized control applies optimization algorithms to solve a centralized voltage regulation problem based on the operation information of the whole system [13]. This is challenging since the optimal power flow (OPF) is a non-convex optimization problem. To this end, heuristic methods [3], approximated approaches [4], non-linear optimization [5] and convexification methods [6] have been developed. To further address the uncertainties of renewable energy generations, the stochastic programming (SP)-based approaches are developed [7]. However, the SP-based ones have a heavy computational burden as many scenarios need to be considered. Also, SP requires the accurate distribution of random variables, which may not be possible in practice. Different from SP, robust optimization (RO) methods deal with the uncertainty by constructing an uncertainty set and obtain the solutions under the worst scenarios [8-10]. As a result, the outcomes are typically conservative. Model predictive control [11-12] is another way of addressing voltage regulation but has difficulties for large-scale systems. In summary, the centralized strategies rely heavily on the communication links and require long calculation time. Its scalability for large-scale systems remains to be an issue.

The distributed autonomous control strategies make decisions for voltage regulation based on local observations [14-15]. They are easy to implement but has the problem of finding the global optimal solution due to the lack of cooperation between various control subjects. By contrast, the distributed cooperative control can achieve the coordination between different units with limited communication links [16]. Among them, the partition-based distributed coordination control is attractive in recent years [10], [17-19]. The main idea is to apply a clustering algorithm to partition the whole network into several sub-networks according to the predefined electrical distance. Then the optimization methods are applied to achieve distributed voltage regulation of each cluster. The optimization methods [10], [17-19] need to pre-determine an optimal solution so that the uncertainties of photovoltaic (PV) outputs and load demand can be mitigated. However, as renewable energy generations depend on the environmental factors, the PV outputs can significantly fluctuate when radiation changes fast. According to [20], the PV output may vary by 15% of its rated power in less than one minute. Under this condition, more frequent operations of multiple devices are needed to provide flexible responses to system uncertainties. However, the aforementioned methods need to resolve the optimization problem when a new situation is encountered, leading to high computational burdens. Additionally, these methods are model-based and require accurate parameters of the DN, which is challenging for practical DN [21].

To mitigate the model quality issue, machine learning (ML)-based voltage regulation methods for DN are developed. ML

methods can extract knowledge from historical data to deal with system uncertainties. The extracted knowledge has generalizability to new situations without resolving the problem, which can provide decisions in real-time [22-23]. Among various ML methods, deep reinforcement learning (DRL) can learn optimal control strategies from data and is suitable for the control and optimization problems [21], [24]. The relationships between the state and action are learned via numerous samples generated by continuous interactions with the environment, thus reducing the dependence on the knowledge of system parameters. However, they are based on the centralized control framework and the distributed coordination of various control devices is missing, leading to suboptimal solutions.

This paper proposes a novel distributed coordination control strategy based on the unsupervised clustering and multi-agent DRL (MADRL) for DN voltage regulation considering reactive power capability of static var compensation (SVC) and PV inverters. The main contributions are:

• The proposed approach is developed based on a novel framework, namely centralized training, and decentralized execution. It enables the agents to exhibit cooperative behaviors based on local information without specific assumptions on the communication channel between sub-regions of the DN. Specifically, the spectral clustering algorithm is used to partition the DN into serval sub-networks according to different levels of coupling relationship between voltage and reactive power. Each sub-network is modeled by a twin delayed deep deterministic policy gradient (TD3) agent. All agents are trained in a centralized manner to learn the coordination control strategy and can be executed in a distributed manner to provide near-optimal decisions based on the latest local information. This helps significantly reduce the cost of deploying additional communication infrastructure.

• The optimization problem of each sub-network is solved by the DRL algorithm. DRL can extract voltage regulation strategy from historical data and inform decisions in real-time according to the latest observed information to reduce voltage deviation. Thus, the proposed approach can effectively mitigate system uncertainties and no pre-determined decisions are needed as compared to other methods.

The rest of the paper is organized as follows. In section II, the problem formulation is presented. Section III describes the proposed method. In section IV, the simulation results are illustrated in detail. Finally, Section V concludes this paper.

## II. PROBLEM FORMULATION

The voltage regulation problem is formulated as follows:

$$\min_{Q_{i,t}^{PV}, Q_{i,t}^{SVC}} F(x) = \sum_{i \in N} \sum_{t=1}^{T} |V_{i,t} - V_0| \quad (1)$$

$$s.t \quad P_{i,t}^{PV} - P_{i,t}^{Load} = V_{i,t} \sum_{j=1}^{N} V_{j,t} (G_{ij,t} \cos\theta_{ij,t} + B_{ij,t} \sin\theta_{ij,t}), \forall i \in N \quad (2)$$

$$Q_{i,t}^{PV} + Q_{i,t}^{SVC} - Q_{i,t}^{Load} = V_{i,t} \sum_{j=1}^{N} V_{j,t} (G_{ij,t} \sin\theta_{ij,t} - B_{ij,t} \cos\theta_{ij,t}), \forall i \in N \quad (3)$$

$$V_{i,\min} \leq V_{i,t} \leq V_{i,\max}, \forall i \in N \quad (4)$$

$$|I_{ij,t}| \leq I_{ij,\max}, \forall (i,j) \in E \quad (5)$$

$$Q_{\min}^{SVC} \leq Q_{i,t}^{SVC} \leq Q_{\max}^{SVC}, \forall i \in N \quad (6)$$

$$0 \leq P_{i,t}^{PV} \leq P_{i,\max}^{PV}, \forall i \in N \quad (7)$$

$$(P_{i,t}^{PV})^2 + (Q_{i,t}^{PV})^2 \leq (S_i^{PV})^2, \forall i \in N \quad (8)$$

where (1) is the objective function to minimize the sum of the voltage deviation of each node; $Q_{i,t}^{PV}$ and $Q_{i,t}^{SVC}$ are control variables, which denote reactive power injection of PV inverter and SVC that is connected to node $i$ during time $t$, respectively; $V_{i,t}$ is the voltage of node $i$ during time $t$; $V_0$ is the rated voltage; $N$ is the set of buses in the entire system; (2)-(3) are the equality power flow constraints, where $P_{i,t}^{PV}$ is the active power injection of PV connected to node $i$ during time $t$; $P_{i,t}^{Load}$ and $Q_{i,t}^{Load}$ are the active and reactive power of load demand connected to node $i$ during time $t$; $G_{ij,t}$ and $B_{ij,t}$ are the real and imaginary part of admittance element between nodes $i$ and $j$; $\theta_{ij,t}$ is the voltage phase difference between nodes $i$ and $j$; (4)-(8) are inequality constraints. Specifically, (4) is the constraint of the voltage for each node, where $V_{i,\min}$ and $V_{i,\max}$ are the lower and upper limits; (5) refers to the branch current limit; $I_{ij,t}$ and $I_{ij,\max}$ represent the current magnitude of branch $ij$ during time $t$ and upper limit of branch current magnitude, respectively; $E$ represents the set of edges in the entire system; (6) describes the reactive power range of SVC, where $Q_{\min}^{SVC}$ and $Q_{\max}^{SVC}$ are the lower and upper limits; (7)-(8) are the constraints of PV, where $P_{i,\max}^{PV}$ and $S_i^{PV}$ are the rated and apparent power of PV connected to node $i$; (7) denotes that the active power of PV should be within the range of rated power; (8) denotes that the reactive power of the PV inverter connected to node $i$ depends on the active power of PV during time $t$.

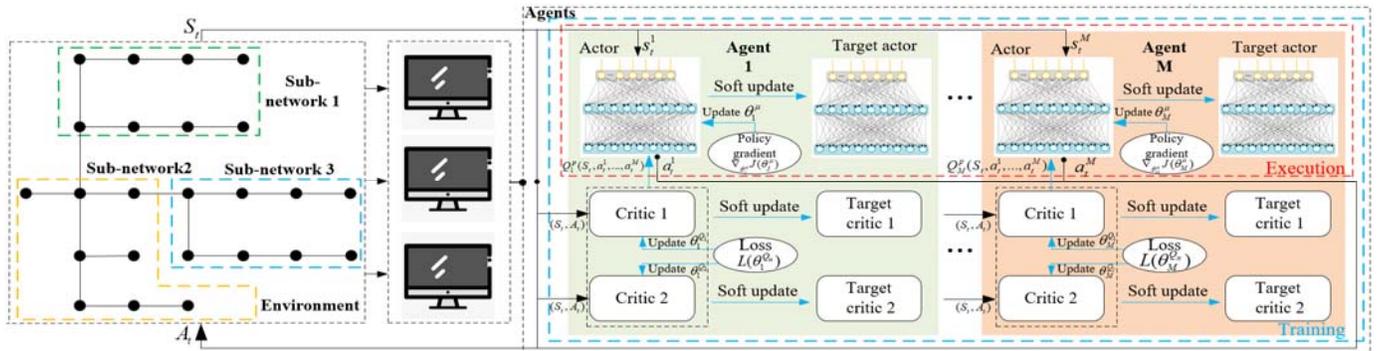

Fig. 1. The architecture of the proposed MATD3-based coordinated control for DN voltage regulation.

## III. PROPOSED DISTRIBUTED MATD3 CONTROL

The proposed distributed multi-agent TD3 (MATD3) algorithm for voltage regulation is shown in Fig. 1. It contains three main components, namely i) the clustering algorithm for decomposing the DN; ii) formulation of the decomposed sub-networks as Markov games and iii) voltage control optimization via the proposed MATD3 algorithm.

### A. Network Partition Method

The objective of clustering is to partition the DN into several sub-networks to identify the voltage control areas. This is similar to the case that the centralized optimization problem is divided into several small problems and is solved in a decentralized manner. Spectral clustering is an unsupervised learning method derived from spectral graph theory. Compared with other clustering algorithms, spectral clustering is more suitable for partition networks with graph structure. Therefore, the spectral clustering algorithm is used to search for the optimal partition results of DN in this paper.

Since this paper aims to reduce voltage deviation utilizing reactive power of multi resources, the voltage-reactive power sensitivity matrix is used to represent the electrical distance. The affinity matrix is derived based on the sensitivity matrix via

$$w_{i,j} = w_{j,i} = \sum_{i=1, j=1}^{N} \exp(\frac{-\|x_i - x_j\|^2}{2\sigma^2}) \quad (9)$$

where $w_{i,j}$ is the component of the similarity matrix; $x_i$ represents the $i$th row of the sensitivity matrix; $\sigma$ is the coefficient that controls the width of the neighborhood. Then the diagonal degree matrix $D$ can be obtained by $d_i = \sum_{j=1}^{n} w_{ij}$, where $d_i$ denotes the $i$th diagonal element of $D$. After that, the Laplacians matrix $L$ is calculated via $L = D - W$.

The clustering problem is transferred to a graph partition problem and the objective function is as follows [25]:

$$F_{Ncut}(A_1, A_2, ..., A_k) = \frac{1}{2} \sum_{i=1}^{k} \sum_{m \in A_i, n \in \bar{A}_i} \frac{w_{m,n}}{vol(A_i)} \quad (10)$$

where $k$ is the total number of clusters; $A_i$ denotes the $i$th group of the clustering results; $\bar{A}_i$ is the complement set of $A_i$; $vol(A_i)$ represents the weighted sum of all edges in the group $A_i$. The objective of (10) is to maximize the internal similarity of subgraphs and minimize the similarity between subgraphs. According to [25], the optimization of (10) can be transferred to:

$$\arg\min tr(F^T D^{-1/2} L D^{-1/2} F) \quad (11)$$
$$s.t. \quad F^T F = I$$

The optimal partition results can be obtained by constructing a space using the eigenvectors corresponding to the first $k_1$ eigenvalues of the matrix $D^{-1/2} L D^{-1/2}$, and clustering the eigenvectors in the space by K-means algorithm.

### B. Formulation of Markov Games

In this section, the voltage regulation of multi sub-networks is formulated as Markov Games (MGs). In the MGs, each sub-network is modeled as an adaptive agent, which makes control decisions based only on local information of corresponding sub-network at each time step. The key components for a MG include state set $S$, action set $A$, and reward function $R$.

- $S$: the state set $S_t$ contains the states for all agents. For agent $j$, the state at time step $t$, $s_t^j$, is the local observation of sub-network $j$, which includes $(P_{i,t}^{Load}, Q_{i,t}^{Load}, P_{i,t}^{PV})$, where $i$ is the index of the node that is located in sub-network $j$.
- $A$: the action set $A_t$ contains the actions for all agents. For agent $j$, the action at time step $t$, $a_t^j$, includes $(Q_{i,t}^{SVC}, Q_{i,t}^{PV})$.
- $R$: $r_t^j \in R_t$ is the immediate reward the agent $j$ obtains after the action $a_t^j$ is executed. In this context, all the agents share the same reward: $r_t = \sum_{i \in N} |V_{i,t} - V_0|$, which represents the total voltage deviation of all nodes in the DN at time step $t$.

At time step $t$, agent $j$ makes a decision $a_t^j$ based on the local observation $s_t^j$ of sub-network $j$. When all agents complete actions, they obtain a shared reward $r_t$, and then the system transfers to the next state. This is an MG and the aim of each agent is to learn a policy, which maps its local observation $s_t^j$ to action $a_t^j$ to maximize the discounted cumulative reward from the current time-step onward, $\sum_{k=t}^{T} \gamma^{k-t} r_k$, where $\gamma \in [0,1]$ is the discount factor that balances the importance between the future and immediate reward.

### C. Proposed MATD3 Algorithm for Voltage Control

The MATD3 algorithm is developed to solve the formulated MGs. To this end, each sub-network is modeled as a TD3 agent, which is composed of the actor and critic. The actor maps the local observation $s_t^j$ to action $a_t^j$ and is the policy function. The critic maps the global information $(S_t, A_t)$ to a scalar, which is a judgment of action $a_t^j$ considering the impact on other agents. The coordinated control strategy is achieved by adopting a centralized training framework, among which the actor and critic functions of each agent are trained against each other iteratively until the critic provides better judgment and the actor can make decisions with reduced voltage deviation.

#### 1) Actor and Critic Functions

The actor parameterized by $\mu_j$ is the policy function and it aims to maximize the output of the critic by making the decision $a_t^j$ under the state $s_t^j$. Let $p = \{p^{\mu_1}, ..., p^{\mu_M}\}$ represents the set of policies for $M$ agents, then the policy function is optimized by directly adjusting the parameters $\mu_j$ toward the direction that maximizes the following objective [26]:

$$\nabla_{\mu_j} J(\mu_j) = E_{S_t, A_t \sim D}[\nabla_{\mu_j} p^{\mu_j}(a_t^j | s_t^j) \nabla_{a_t^j} Q_j^p(S_t, a_t^1, ..., a_t^M)|_{a_t^j = p^{\mu_j}(s_t^j)}] \quad (12)$$

where $Q_j^p(S_t, a_t^1, ..., a_t^M)$ is the centralized action value function and is approximated by the critic function.

The critic function parameterized by $Q_j$ takes the global state and actions of all agents as inputs and outputs the action value of agent $j$. The critic function is optimized by minimizing

$$L = (Q_j^p(S_t, a_t^1, ..., a_t^M) - y)^2 \quad (13)$$

$$y = r_t^j + \gamma Q_j^p(S_{t+1}, a_{t+1}^1, ..., a_{t+1}^M)\big|_{a_{t+1}^i = p^{\mu_i}(s_{t+1}^i)} \quad (14)$$

where $Q_j^p(\cdot)$ is the output of the critic function; $y$ is the target. The critic function is optimized by minimizing the "distance" between $Q_j^p(S_t, a_t^1, ..., a_t^M)$ and the target. However, the training process is unstable in many environments since the critic being updated is also used for calculating the target $y$. To solve this problem, the target functions $Q_j'^{p'}(S_{t+1}, a_{t+1}^1, ..., a_{t+1}^M)$ and $p^{\mu_j'}$ are introduced. $y$ is calculated by target functions, the parameters of which are tracking the online functions: $\mu_j' \leftarrow \tau\mu_j + (1-\tau)\mu_j', Q_j' \leftarrow \tau Q_j + (1-\tau)Q_j'$. It is able to significantly improve the stability of the training process. Then (14) can be rewritten as:

$$y = r_t^j + \gamma Q_j'^{p'}(S_{t+1}, a_{t+1}^1, ..., a_{t+1}^M)\big|_{a_{t+1}^i = p^{\mu_i}(s_{t+1}^i)} \quad (15)$$

The overestimated value caused by the function approximation error in the actor-critic based method usually leads to suboptimal policies. To address that, TD3 algorithm mimics the mechanism of Double Q-learning by utilizing a pair of critics $(Q_{j,1}^p, Q_{j,2}^p)$ to calculate the target $y$ [27]:

$$y = r_t^j + \gamma \min_{n=1,2} Q_{j,n}'^{p'}(S_{t+1}, a_{t+1}'^1, ..., a_{t+1}'^M)\big|_{a_{t+1}^i = p^{\mu_i}(s_{t+1}^i)} \quad (16)$$

To further address the overfitting issue in deterministic DRL methods, a regularization term is introduced to smooth the target policy. Thus, the target action $a_{t+1}'^i$ in (16) is replaced by

$$\tilde{a}_{t+1}^i = a_{t+1}'^i + \varepsilon \quad (17)$$

$$\varepsilon \sim clip(N(0,\sigma), -c, c) \quad (18)$$

where $clip(.)$ is used to clip the added noise to keep the target close to the original action; $-c$ and $c$ represent the lower and upper bounds of the added noise, respectively.

*2) DNN for Function Approximation*

To handle the uncertainties and dynamics of DN as well as the nonlinearity when solving power flow, DNN is advocated to approximate the actor and critic functions in the TD3 algorithm [28]. For each DNN, the relationship between input $x$ and output $y$ can be expressed as:

$$y = f_l[...f_1[x]] \quad (19)$$

$$o_i = f_i[o_{i-1}] = g(W_i * o_{i-1} + b_i) \quad (20)$$

where $f_i(\cdot)$ is the function map of the $i$th layer of DNN; $o_i$ is the output of the $i$th layer; $W_i$ and $b_i$ are the weight matrix and bias of the $i$th layer, respectively; $g(\cdot)$ is the activation function. Then parameters $\mu_j$ and $Q_j$ are replaced by the parameters of DNN. In the MGs with $M$ agents, the parameter set is $\theta = \{\theta_1, ..., \theta_M\}$. For agent $j$, the parameter set is denoted as $\theta_j = \{\theta_j^\mu, \theta_j^{\mu'}, \theta_j^{Q_1}, \theta_j^{Q_1'}, \theta_j^{Q_2}, \theta_j^{Q_2'}\}$, where $\theta_j^\mu$ and $\theta_j^{\mu'}$ are parameters of actor and target networks of agent $j$; $\theta_j^{Q_1}, \theta_j^{Q_1'}, \theta_j^{Q_2}$ and $\theta_j^{Q_2'}$ are parameters of critic and target critic networks $Q_1$ and $Q_2$, respectively. For each network, the parameter set is $\{W_1, b_1, ..., W_l, b_l\}$.

*3) Optimization of the Actor and Critic Networks*

TABLE I
Training Process of the Proposed Algorithm

| **Algorithm** Training the proposed algorithm |
|---|
| **Input**: the node active and reactive power in the DN, the PV output, and the reward. |
| **Output**: DNN's parameters $\theta$ |
| 1: Randomly initialize parameters of critic networks $\theta_j^{Q_1}$, $\theta_j^{Q_2}$ and actor network $\theta_j^\mu$ for each agent $j$ |
| 2: Initialize target networks $\theta_j^{Q_1'} \leftarrow \theta_j^{Q_1}, \theta_j^{Q_2'} \leftarrow \theta_j^{Q_2}, \theta_j^{\mu'} \leftarrow \theta_j^\mu$ for each agent $j$ |
| 3: for episode =1, 2, ..., H do |
|     receive initial observation $s_0^j$ for each agent $j$ |
|     for $t=1, 2, ...T$ do |
| 4:     choose action according to $a_t^j = p^{\theta_j^\mu}(s_t^j)$ for each agent $j$ |
|     execute actions $A_t = (a_t^1, ..., a_t^M)$ and obtain reward $r_t^j$ and new observation $s_{t+1}^j$ for each agent $j$ |
| 5:     store transition $(S_t, A_t, R_t, S_{t+1})$ in the replay buffer |
|     for agent $j = 1,..., M$ do |
| 7:     sample a random mini-batch $B$ of transitions from replay buffer |
| 8:     calculate target $y$ according to (16) |
| 9:     update critic networks according to (21) |
|     if $t$ mod $d$ do |
|     update actor network according to (23) |
| 10:     update target networks for each agent |
|     end if |
|     end for |
| 11: end for |
| 12: end for |

In the proposed approach, each agent has a replay buffer, which is in charge of storing the transitions $(s_t^j, a_t^j, r_t^j, s_{t+1}^j)$. The mini-batch experiences are sampled at each time step to calculate the gradient and optimize the parameters of networks. This mechanism helps break the correlation between data and improves the stability of the training process. Since the replay buffer mechanism is introduced, the optimization function should be rewritten. Supposing that the mini-batch of experience $(S_t, A_t, R_t, S_{t+1})_k, k=1,2,...,B$ is sampled, then the parameters of critic networks are updated by minimizing:

$$L(\theta_j^{Q_n}) = \frac{1}{B}\sum_{k=1}^{B}(Q_{j,n}^p(S_t, a_t^1, ..., a_t^M) - y)^2, \quad n=1,2 \quad (21)$$

where target $y$ is calculated by (16). The gradient descent is updated as:

$$\theta_j^{Q_n} \leftarrow \theta_j^{Q_n} + \eta_Q \nabla_{\theta_j^{Q_n}} L(\theta_j^{Q_n}), \quad n=1,2 \quad (22)$$

where $\eta_Q$ is the learning rate for the critic function. The gradient of the actor networks is calculated via

$$\nabla_{\theta_j^\mu} J(\theta_j^\mu) = \frac{1}{B}\sum_{k=1}^{B}\nabla_{\theta_j^\mu} p^{\theta_j^\mu}(a_t^j | s_t^j)\nabla_{a_t^j} Q_j^p(S_t, a_t^1,...,a_t^M)\big|_{a_t^j = p^{\theta_j^\mu}(s_t^j)}$$
(23)

Then, the parameters of actor network are through
$$\theta_j^\mu \leftarrow \theta_j^\mu + \eta_\mu \nabla_{\theta_j^\mu} J(\theta_j^\mu) \quad (24)$$

where $\eta_\mu$ is the learning rate for actor function.

### D. Centralized Training and Decentralized Implementation

The parameter set of the proposed approach is $\theta = \{\theta_1,...,\theta_M\}$, where $\theta_j$ denotes the parameter set of agent $j$. In the proposed approach, all agents are trained in a centralized manner to learn a coordinated control strategy and the training process is shown in Table I.

TABLE II
Real-time Reactive Power Control of the Proposed Approach

| Algorithm Real-time reactive power control |
|---|
| **Input**: the node active and reactive power in the DN, the PV output. |
| **Output**: reactive power schedules: $A_1 : A_T$. |
| 1: Read the parameters of actor network of each agent $\theta_j^\mu$ |
| 2: For time step $t$=1, 2, …$T$ do |
| 3:     for agent $j = 1, …, M$ do |
| 4:         obtain the local observation $s_t^j$ |
| 5:         calculate action $a_t^j$ according to $a_t^j = p^{\theta_j^\mu}(s_t^j)$ |
| 6:     end for |
| 7:     concatenate actions of all agents $A_t = (a_t^1,...,a_t^M)$ |
| 8: end for |
| 9: **Return**: $A_1 : A_T$ |

When the training process is completed, the parameters of DNN are fixed and only the actor network of each agent is kept for real-time voltage regulation. Each agent is in charge of a sub-network. The real-time reactive power control scheme of the proposed approach is shown in Table II. The centralized critics that are augmented with information about other agents' policies during training help the formulation of coordinated strategies. The explicitly modeling of other agents' decision-making process allows each agent to provide decisions with better robustness to system dynamics based on local information only. This differentiates the existing works and allows us to deal with scalability issues in the presence of large-scale systems.

## IV. NUMERICAL RESULTS

In this section, simulation results are provided to evaluate the performance of the proposed approach on IEEE 33-bus and 123-bus systems. The network partition results are first illustrated followed by the control performance comparisons results with other methods.

### A. Simulation Setup

To simulate more realistic scenarios, real-world PV data are used, i.e., one-year PV generation data of Xiaojin, a county in Sichuan province of China. These data are divided into a training set and a test set, which contain 300- and 30-days' data, respectively. The parameters of the control devices are shown in Table III. The maximum voltage deviation is set to $\pm 5\%$. For the proposed method, each sub-region is modeled as an agent, which is composed of actor and critic networks. All the networks have two shallow layers, the number of neurons of which are 100 and 100, respectively. The hyper-parameters are shown in Table IV. The proposed approach is implemented in Python with TensorFlow. A workstation with an NVIDIA GeForce 1080Ti GPU and an Intel Xeon E5-2630 CPU is used for the simulation.

Table III Parameters of Control Devices for 33-bus System

| Type | Capacity | Location |
|---|---|---|
| SVC | 0.3MVar | 7, 16, 31 |
| PV | 0.8MWh/0.84MVA | 14, 18, 22, 24, 29, 33 |

Table IV Parameter Settings of the Proposed Method

| Parameters | Values |
|---|---|
| Batch size for updating NN | 32 |
| Replay buffer size | 48000 |
| Discount factor | 0 |
| Soft update coefficient | 0.001 |
| Policy update frequency | 2 |
| Target policy smoothing coefficient | 0.2 |
| Learning rate for actor network | 0.001 |
| Learning rate for critic network | 0.002 |

### B. Network Partition Results

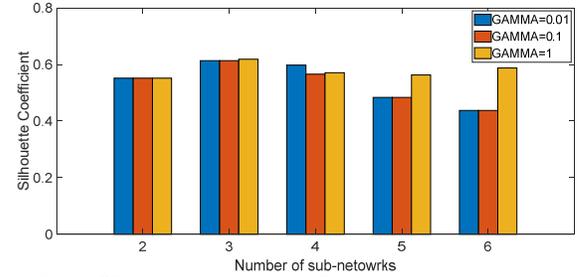

Fig. 2. The Silhouette score with a different number of sub-networks.

The spectral clustering algorithm is applied to partition the DN into several sub-regions. The Silhouette score [29] is utilized as the index to evaluate the performance of the clustering results. The value of the Silhouette score ranges from -1 to 1. A higher value of the Silhouette score indicates a closer electrical connection inside the sub-region and a looser connection among various sub-regions. The results of the Silhouette score with a different number of sub-networks and $\gamma$ values, i. e., 0.01, 0.1, and 1, are shown in Fig. 2. It can be observed from the figure that when the number of sub-networks is 3, the Silhouette score always achieves the highest value. This demonstrates that when the number of sub-networks is set to 3, the connection of nodes inside the sub-networks is the closest. Therefore, the number of sub-networks is selected as 3 for the 33-bus system. The clustering result is shown in Fig. 3. Similarly, the number of sub-networks for the IEEE 123-bus system is chosen to be 5, see Fig. 4.

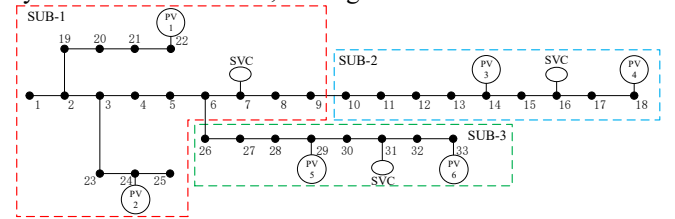

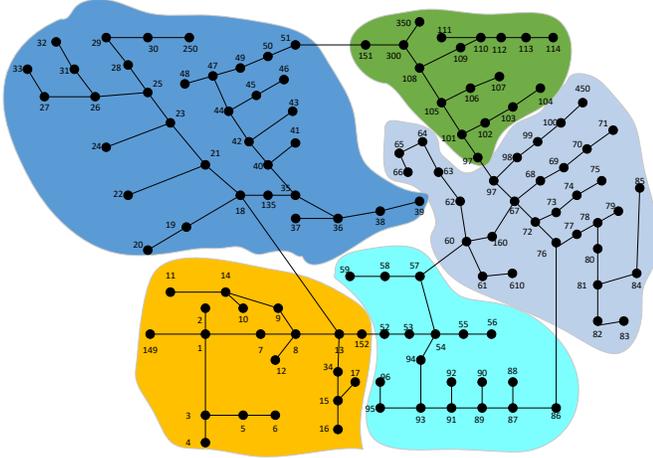

Fig. 3. The partition results of IEEE 33-bus system.

Fig. 4. The partition results of the IEEE 123-bus system.

*C. Performance Evaluation*

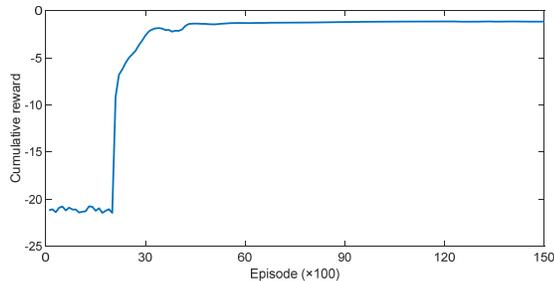

Fig. 5. The evolution of the reward during the training procedure.

The proposed approach is trained for 15000 epochs on the training data to learn the coordinated control strategy for voltage regulation. Each epoch corresponds to a day. The cumulative reward of each epoch is calculated by $\sum_{t=1}^{T} r_t$. The convergence curve of the cumulative reward is plotted in Fig. 5. It can be observed that the proposed approach could not make balanced decisions at the beginning of the training procedure and therefore achieves low reward. With the training process going on, the reward increases significantly and finally converges around -1.3 with small fluctuations. This illustrates that the proposed method can learn the coordinated control strategy from training data.

To assess the generalization ability of the learned strategy from training data and demonstrate the effectiveness of the proposed method, comparative tests are carried out against various existing methods. They include 1) the original method without control; 2) the stochastic programming-based (SP) approach, where all the sub-networks are optimized separately, and the objective of each sub-network is to minimize the voltage deviation based on local information. 300 scenarios are randomly generated to represent the uncertainty and scenarios reduction is used to obtain 20 representative scenarios [7]; 3) the TD3 method, where each sub-network is controlled by a TD3 agent based on the local observation. The TD3 agents are trained separately to minimize the voltage deviation of its own sub-network. Note that there is no information exchange between TD3 agents during the training process. The neural network structure and the hyper-parameters settings of the TD3 agents are the same as the proposed method; 4) the model-based centralized control method, where the commercial SOCP solver MOSEK is applied to solve deterministic cases based on the global information. Its results are considered as benchmarks.

TABLE V
Voltage Deviation of Various Methods on Test Data

| Method | Average | Max rise | Max drop | Time (s) |
|---|---|---|---|---|
| Original | 1.72% | 5.89% | 7.11% | - |
| SP | 0.61% | 3.05% | 1.62% | 140 |
| TD3 | 0.57% | 3.03% | 1.54% | 0.0012 |
| Proposed | 0.17% | 0.68% | 1.25% | 0.0012 |
| Centralized | 0.11% | 0.62% | 1.21% | 0.84 |

The average, maximum rise, and maximum drop of voltage deviations, as well as the computing times for all methods, are shown in Table V. It can be observed that when reactive compensation is not applied, the voltage will exceed the upper and lower limits. The SP, TD3, and the proposed approach can ensure the voltage to be within the limited ranges. Note that these three methods all make decisions based on local information. However, the proposed approach achieves much better performance due to the coordinated control strategies learned during the training process. To evaluate the control accuracy of the proposed method, the average optimization error is defined:

$$ERR = |\frac{\Delta V_{pro} - \Delta V_{cen}}{\Delta V_{cen} - \Delta V_{ori}}| \times 100\% \quad (25)$$

where *ERR* represents the average deviation of the proposed approach to the global optimal solution; $\Delta V_{cen}$, $\Delta V_{pro}$ and $\Delta V_{ori}$ represent the average voltage deviations of the centralized approach, the proposed approach, and the original value on test data, respectively. The *ERR* is 3.7% for the proposed approach and this means that it can reach 96.3% optimality based on local information. This demonstrates the effectiveness of the proposed partition-based control strategy. It can also be found from Table V that the proposed approach and TD3 method take much fewer calculation times than other ones. This is because the control strategy learned by the proposed approach is scalable and can be generalized to newly encountered situations. This process is similar to recalling from the memory without resolving the problem again, yielding high computational efficiency.

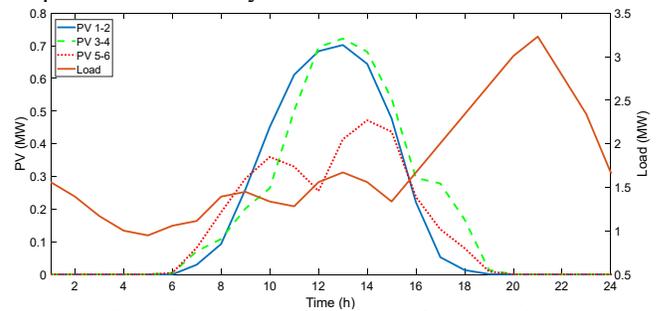

Fig. 6. PV outputs and load demand for a sunny day.

To further demonstrate the capability of the proposed method in dealing with fluctuations of PV and load outputs, a sunny day is selected as a case study. The PV output and load demand for a sunny day are shown in Fig. 6. The voltage profile of each node under various control strategy is shown in Fig. 7 when *t=1:00 PM*. It can be observed that the voltages at nodes 15-18 go beyond the upper voltage limit when there is no reactive compensation. With the SP and TD3 methods, those voltages can be adjusted to a limited range. However, compared with the proposed approach, they suffer from over-adjustment due to the lack of coordination of multi-devices. Since the proposed approach explicitly models the decision-making process of other agents via centralized training, the agents can exhibit cooperative behaviors using only local information for voltage regulation. The voltage profile of the proposed approach can be found to be very close to that of the centralized method, demonstrating its capability of reaching the global optimal solutions. The voltage profiles of node 17 for various approaches across a whole day are also shown in Fig. 8. The test results are consistent with those observed from Fig. 7.

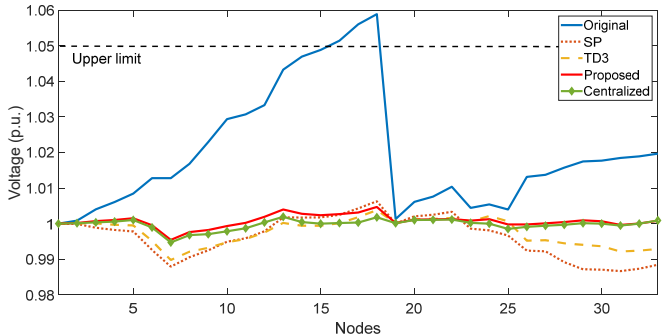
Fig. 7. Voltages of each node before and after control when *t=1:00* PM.

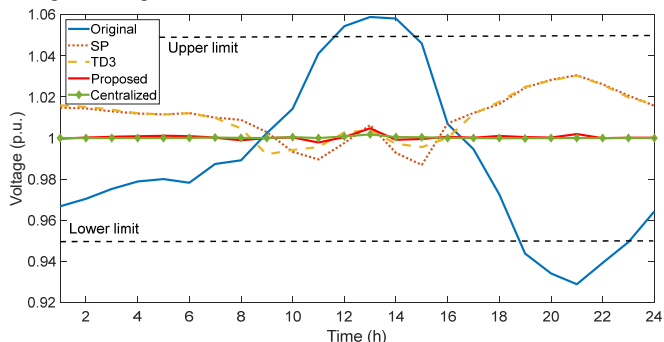
Fig. 8. Voltage change of node 17 before and after control.

### D. Robustness to Large Stochasticity

More simulations are carried out to demonstrate the advantage of the proposed approach for real-time controls. The PV output profile is shown in Fig. 9. Due to the cloud dynamics, the PV output changes fast within 1 minute, i.e., its output rises from 0.33 MW at *t=1s*, reaches 0.65 MW at *t=30s*, and then returns to 0.33 MW in 30 seconds. The voltage profiles of node 18 under various control strategies are shown in Fig. 10. For the SP method, a predetermined control solution is used to react to the uncertainty of the PV output while the TD3 method and the proposed approach provide decisions at each second. The centralized method ignores the communication delay and provides a theoretical limit for the problem (an ideal condition that could not be achieved in practice). It can be observed from Fig. 10 that node 18 suffers from an overvoltage problem if no controls of SVC and PV inverter are applied, namely the original method. With the SP method, the problem can be suppressed. However, since the control decisions provided by this approach are predetermined and cannot react dynamically to the fast-changing PV outputs, it has much larger voltage fluctuations than those of the TD3 method and the proposed method. By contrast, the TD3 method and the proposed approach can make decisions based on the latest states of the DN, and thus can achieve a better response to the dynamic changes of the PV outputs. The proposed approach further outperforms TD3 due to the coordinated control strategy learned during the training process.

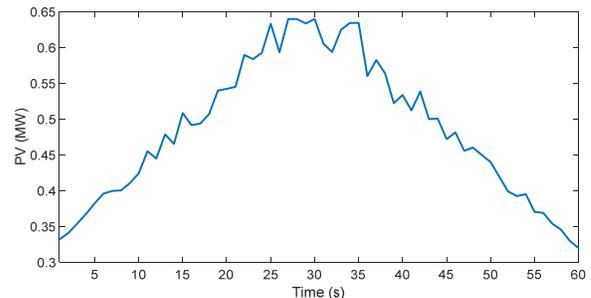
Fig. 9. PV outputs with large variations.

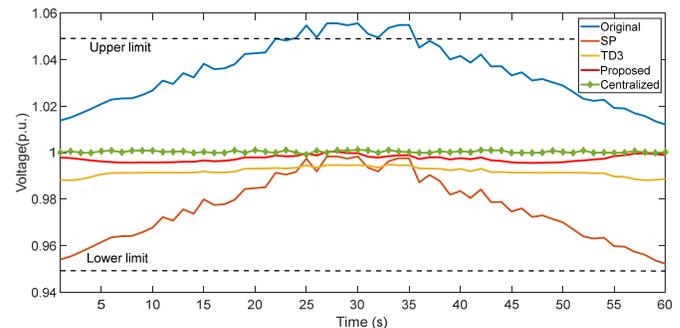
Fig. 10. Voltage change of node 18 with different control strategies when the PV outputs have large fluctuations.

### E. Scalability to IEEE 123-bus System

Table VI
Control Devices in 123-bus System

| Type | Capacity | Location |
|---|---|---|
| SVC | 0.3MVar | 7, 46, 50,77,92,99,109 |
| PV | 0.8MWh/0.84MVA | 4,9,24,31,56,59,63,70,75,106 |

TABLE VII
Voltage Deviations obtained from Various Control Methods for IEEE 123-bus System

| Method | Average | Max rise | Max drop | Calculation time (s) |
|---|---|---|---|---|
| Original | 1.73% | 6.29% | 4.33% | - |
| SP | 3.1% | 4.77% | 8.31% | 430 |
| TD3 | 2.93% | 4.59% | 7.9% | 0.0012 |
| Proposed | 0.46% | 1.65% | 2.9% | 0.0012 |
| Centralized | 0.32% | 1.33% | 2.7% | 1.83 |

To assess the scalability of the proposed method to a larger-scale system, tests are carried out on the IEEE 123-bus system. The parameter settings of the control devices are shown in Table VI, including the capacities and locations of SVCs and PVs. According to the network partition criterion developed in Section III-A, it is found that the Silhouette score obtains the highest value of 0.872 when the clustering number is set to 5. The clustering result is shown in Fig. 4 and that means there are five agents, each corresponding to a sub-network. The parameter settings of the proposed control method are the same as those for the IEEE 33-bus system. The voltage deviations obtained from different approaches to test data are shown in Table VII. It can be found that if there is no control implemented, the maximum voltage rise will be 6.29%. It is very interesting to find that both SP and TD3 have serious voltage control issues for the 123-bus system. The SP failed to find a good voltage regulation strategy based on local information. The identified solution has a maximum 8.31% voltage drop, which is even worse for the system without control. For the TD3 methods, with the increased size of the system and the problem complexity, the negative impacts of not coordinating with different sub-networks have been shown here. The control based on only local information is not sufficient. By contrast, for the small-scale 33-bus system, even without the coordination, TDs can achieve reasonable performance. Nevertheless, the proposed method is implemented with centralized training and distributed execution. The interactions among different agents have been explicitly considered and this allows us to achieve near-global optimal solutions as the centralized optimization.

## V. CONCLUSIONS

This paper proposes data-driven distributed coordination control for distribution system voltage regulation considering PV inverters and SVCs. The spectral clustering algorithm allows us to partition the large distribution system into several sub-networks from the voltage control perspective. Then, the control of each sub-network is formulated as the MGs and solved by the MATD3 algorithm. The proposed method is built based on the centralized training distributed implementation framework and can be easily used for real-time voltage regulation. Comparative results with several other existing model-based and data-driven methods demonstrate that the proposed method can achieve 96.3% optimality based on local information while considering the uncertainty. However, the model-based methods have difficulties in dealing with uncertainties while other data-driven could not achieve satisfactory outcomes in the presence of large variations of PV outputs. Future work will be on testing the developed method using realistic systems with field data.